\documentclass[12pt]{article}
\newcommand{\dd}{\mathrm{d}}
\newcommand{\dfrac}[2]{\frac{\displaystyle #1}{\displaystyle #2}}
\begin{document}
\title{\normalsize \hfill UWThPh-2011-34 \\[1cm] \LARGE
On the 100th anniversary of the \\ Sackur--Tetrode equation 
}
\author{W. Grimus\thanks{E-mail: walter.grimus@univie.ac.at} 
\setcounter{footnote}{6} 
\\[4mm]
\small University of Vienna, Faculty of Physics \\
\small Boltzmanngasse 5, A--1090 Vienna, Austria \\[4mm]
\\[5mm]}

\date{23 January 2013}

\maketitle

\begin{abstract}
In 1912, Otto Sackur and Hugo Tetrode independently put forward an equation for
the absolute entropy of a monoatomic ideal gas and published it in
``Annalen der Physik.'' The grand achievement in the
derivation of this equation was the discretization of phase space for massive
particles, expressed as $\delta q \delta p = h$, where $q$ and $p$ are
conjugate variables and $h$ is Planck's constant. 
Due to the dependence of the absolute entropy on Planck's constant,
Sackur and Tetrode were able to devise a test of their equation by
applying it to the monoatomic vapor of mercury; from the satisfactory
numerical comparison of $h$ obtained from thermodynamic data on mercury 
with Planck's value from black-body radiation, they inferred the
correctness of their equation. In this review we highlight this almost
forgotten episode of physics, discuss the arguments 
leading to the derivation of the Sackur--Tetrode equation and outline
the method how this equation was tested with thermodynamic data. 
\\[1mm]
PACS: 05.70.-a, 51.30.+i
\end{abstract}

\setcounter{footnote}{0}
\renewcommand{\thefootnote}{\arabic{footnote}}

\newpage

\section{Introduction}

The formula for the absolute entropy of a monoatomic ideal gas 
is named after Otto Sackur and Hugo Tetrode
who independently derived it in~1912~\cite{sackur2,tetrode,sackur}. 
In classical thermodynamics the entropy of a monoatomic ideal gas is 
\begin{equation}\label{Skl}
S(E,V,N) = kN \left( \frac{3}{2} \ln \frac{E}{N} + \ln \frac{V}{N} + 
s_0 \right),
\end{equation}
where $E$, $V$ and $N$ are the kinetic energy, the volume and the number of
atoms, respectively. In classical physics the constant $s_0$ is undetermined. 
The achievement of Sackur and Tetrode was to compute $s_0$. 
At first sight this does not look very exciting,
however, in order to compute $s_0$ they had to work out the size of
``elementary cells or domains'' 
in phase space. Only with this knowledge it is possible to count the number of
states in classical phase space which is a prerequisite for the computation
of Boltzmann's absolute entropy given by~\cite{boltzmann,planck3} 
\begin{equation}\label{SW}
S = k \ln W.
\end{equation}
In this formula, $W$ is the number of possibilities to realize a system
compatible with some given boundary conditions. Sackur and Tetrode determined
the volume of phase space cells as $h^n$ where $h$ is Planck's constant and
$n$ is the number of degrees of freedom. Until then, $h$ was primarily
associated with harmonic oscillators and photons. 
With the work of Sackur and Tetrode it became clear that Planck's constant
was not only relevant for counting the number of states in the case
of photons but also in the case of massive particles. 
In this way, $h$ became ubiquitous in
statistical physics, more than ten years before the advent of quantum mechanics.

This was an amazing result because a priori Planck's constant in the
expression $h \nu$ for the energy of a photon has nothing to do 
with the phase-space volume associated with massive particles. This
connection was clarified only later by quantum mechanics. 
We want to stress that the elegance of the work of Sackur and Tetrode
derives from the combination of theoretical considerations and usage
of experimental data with which they were able to
lend credibility to their result. They did so by successfully applying their
equation to the then available data on mercury, whose vapor is monoatomic
and behaves in good approximation as an ideal gas, 

Below we list the articles of Sackur and Tetrode and the achievements therein, 
written in the course of the development of their equation. 
The titles are literal translations from the German titles.
\begin{enumerate}
\renewcommand{\theenumi}{\roman{enumi}}
\item
O.~Sackur,
\textit{The application of the kinetic theory of gases to chemical
  problems}~\cite{sackur1} (received October 6, 1911): In this paper Sackur
develops the formula for the entropy~$S$ of a monoatomic ideal gas as a
function of the size of the elementary cell.
\item
O.~Sackur,
\textit{The meaning of the elementary quantum of action for gas theory and the
computation of the chemical constant}~\cite{sackur2} (no ``received date'',
must have been written in spring 1912): Here Sackur postulates that the size
of the elementary cell is $h^n$ and obtains the absolute entropy $S$ of a
monoatomic ideal gas. Using $S$, he computes the vapor pressure over a solid and
makes a comparison with data on neon and argon. The numerical results,
are, however, not completely satisfying.
\item
H.~Tetrode, 
\textit{The chemical constant and the elementary quantum of
  action}~\cite{tetrode} (received March 18, 1912):
Tetrode gives an illuminating derivation of $S$, assuming that the size of the
elementary cell is $(zh)^n$. He fits the parameter $z$ by using data on the
vapor pressure of liquid mercury. Due to some numerical mistakes he obtains
$z \approx 0.07$.\footnote{Actually, from Tetrode's equations~(12)
  and~(13) we would rather deduce $z \approx 0.02$.} 
\item
H.~Tetrode, erratum to 
\textit{The chemical constant and the elementary quantum of
  action}~\cite{tetrode} (received July 17, 1912): Tetrode corrects the
numerics and obtains now $z \sim 1$. He acknowledges the
papers~\cite{sackur2,sackur1} of Sackur by noting that the formula for
$S$ has been developed by both of them at the same time. More precisely, he
refers to a formula for the so-called ``chemical constant'' pioneered by
Nernst~\cite{nernst}, which we will define later.
\item
O.~Sackur,
\textit{The universal meaning of the so-called elementary quantum of
  action}~\cite{sackur} (received October 19, 1912): 
He obtains good agreement ($\pm 30\%$) with the data on the vapor
pressure of mercury and comments on the paper by
Tetrode. 
\end{enumerate}

The paper is organized as follows. In section~\ref{derivation} we describe
the different approaches of Sackur and Tetrode to derive their equation and
add some comments. Since historically  
the corro\-boration of the Sackur--Tetrode equation 
by using data on the vapor pressure of (liquid) mercury was crucial,
we give a detailed account of it in 
section~\ref{vapor pressure}. Moreover, we redo the numerics by using modern
mercury data in section~\ref{fit} and obtain a reasonably good value
of Planck's constant. In section~\ref{conclusions} our conclusions are
presented. 
A derivation of Kirchhoff's equation, which is used in the numerical
computation, is found in the appendix.

\section{The Sackur--Tetrode equation}
\label{derivation}

\subsection{Tetrode's derivation}
The starting point of Tetrode's reasoning is the entropy formula~(\ref{SW})
which should, according to Nernst's heat theorem~\cite{nernst}, 
give the correct value of the entropy without any
additive constant. Then he considers a system
with $n$ degrees of freedom and phase space coordinates 
$q_1, \ldots, p_n$, for which he 
connects $W$ with the number of configurations of phase
space points. In order to have a finite entropy, it is necessary to discretize
phase space, which Tetrode does by introducing ``elementary domains'' of
volume 
\begin{equation}
\delta q_1\, \delta p_1 \cdots \delta q_n\, \delta p_n = \sigma = (zh)^n,
\end{equation}
where $h$ is Planck's constant and $z$ is a dimensionless number.
Then he argues that, in a system of $\nu$ identical particles, configurations
which are related only by exchange of particles should not be counted as
different. Therefore, denoting by $W'$ the number of
configurations in phase space, the entropy for such a system is
\begin{equation}
S = k \ln \frac{W'}{\nu !}.
\end{equation}
This is to avoid the Gibbs paradox and to obtain $S$ as an extensive quantity,
though Tetrode does not mention Gibbs in this context. 
Moving on to the monoatomic gas consisting of $\nu \equiv N$ atoms with mass
$m$ and spatial volume $V$, the number of degrees of freedom is $n = 3N$ and, 
for a given maximal energy $E$ of the gas, 
the volume occupied in phase space is computed by 
\begin{equation}
\mathcal{V}(E,V,N) = \int \dd^3 x_1 \int \dd^3 p_1 \cdots 
\int \dd^3 x_N \int \dd^3 p_N 
\quad \mbox{with} \quad 
\frac{1}{2m} \left( {\vec p_1}^{\,2} + \cdots + 
{\vec p_N}^{\,2} \right) \leq E.
\end{equation}
Utilizing the gamma function, this phase space volume is expressed as 
\begin{equation}
\mathcal{V}(E,V,N) = \frac{(2\pi mE)^{\frac{3N}{2}}\, V^N}%
{\Gamma\left(\frac{3N}{2} + 1 \right)}.
\end{equation}
According to the arguments above, the entropy is then given by
\begin{equation}\label{S1}
S = k \ln \frac{\mathcal{V}(E,V,N)}{(zh)^{3N} N!}. 
\end{equation}
In the last step Stirling's formula is used, to wit the approximations 
\begin{equation}
\ln N! \simeq N ( \ln N - 1 ) 
\quad \mbox{and} \quad 
\ln \Gamma\left( \frac{3N}{2} + 1 \right) \simeq 
\frac{3N}{2} \left( \ln \frac{3N}{2} - 1 \right)
\end{equation}
for large $N$. This leads to Tetrode's final result 
\begin{equation}\label{s-tetrode}
S(E,V,N) = kN \left( \frac{3}{2} \ln \frac{E}{N} + 
\ln \frac{V}{N} + \frac{3}{2} \ln \frac{4\pi m}{3 (zh)^2} + 
\frac{5}{2} \right)
\end{equation}
for the entropy of a monoatomic ideal gas. 

This derivation is of an amazing lucidity. No wonder that 
100 years later it is one of the standard methods in modern textbooks.
The only amendment to Tetrode's derivation comes from quantum mechanics which
fixes the size of the elementary domain to $h^n$, i.e. requires $z=1$;
the latter result was obtained by Tetrode through a fit to the data of
the vapor pressure of mercury. 

From equation~(\ref{s-tetrode}) with $z=1$ we infer that the constant
$s_0$ of equation~(\ref{Skl}) is given by
\begin{equation}
s_0 = \frac{3}{2}\, \ln \frac{4\pi m}{3h^2} + \frac{5}{2}.
\end{equation}

\subsection{Sackur's derivation}
It is much harder to follow Sackur's line of thoughts. 
Here we sketch the derivation of the entropy formula in~\cite{sackur},
because there he gives the most detailed account of his derivation.
In this paper he first derives Planck's law of radiation
by considering a system of radiators, before he moves on to the ideal
monoatomic gas. In both cases Sackur defines a time interval $\tau$ in which
the system is monitored and an energy interval $\Delta \varepsilon$ for the
discretization of energy. For the gas the time $\tau$ is assumed to be
so small that during this time collisions between atoms can be neglected. 
Therefore, during the time interval of length $\tau$, each of  
the kinetic energies associated with the three directions in space, 
$\varepsilon_x$, $\varepsilon_y$, $\varepsilon_z$, of every atom can be assumed
each to lie in a well-defined energy interval of length $\Delta \varepsilon$. 
In other words, Sackur imagines a three-dimensional energy space with $x$,
$y$ and $z$-axis referring to the kinetic energies 
$\varepsilon_x$, $\varepsilon_y$ and $\varepsilon_z$, respectively,
and with energy unit $\Delta \varepsilon$ on every axis. 
In this way, the energy space is divided into cubes 
of volume $(\Delta \varepsilon)^3$ and the kinetic energy of every
particle lies, during the time interval $\tau$, in a well-defined cube.
If the $i$-th energy cube is given by 
$n_k \Delta\varepsilon \leq \varepsilon_k < (n_k + 1) \Delta\varepsilon$
($k=x,y,z$) with integers $n_k$, the energy $\varepsilon_i$ associated with
this cube can, for instance, be defined as 
\begin{equation}
\varepsilon_i = (n_x + n_y + n_z) \Delta\varepsilon. 
\end{equation}
Sackur considers further the probability $w$ of of observing, during the time
interval $\tau$, atoms with kinetic energy $\varepsilon_k$ ($k=x,y,z$) lying
in a specific energy interval associated with the $k$-axis; he argues that $w$
will be proportional to the product $\tau \Delta \varepsilon$, because the
smaller $\tau$ and $\Delta \varepsilon$ are, the smaller $w$ will be.
Hence, since there are three directions in space, 
the number of atoms in the $i$-th energy cube, $N_i$, will be proportional to 
$(\tau \Delta \varepsilon)^3$. In this way, 
Sackur justifies the Ansatz 
\begin{equation}\label{ansatz}
N_i = N f(\varepsilon_i) \left( \tau \Delta \varepsilon \right)^3,
\end{equation}
where $N$ is the total number of atoms in the volume $V$.

He goes on by distributing the $N$ atoms into $r$ energy cubes, in exactly the
same way as in the case of harmonic oscillators and photons. The number of
possibilities for putting $N_1$ atoms into cube~1, $N_2$ atoms into cube~2,
etc.\ is given by 
\begin{equation}\label{WS}
W = \frac{N!}{N_1! N_2! \cdots N_r!}
\quad \mbox{with} \quad
N = N_1 + N_2 + \cdots + N_r.
\end{equation}
Note that Sackur computes the number of
possibilities $W$ for a given decomposition of $N$ into the
numbers $N_1, \ldots, N_r$, which clearly 
implies that he assumes \emph{distinguishable} atoms; 
for indistinguishable atoms, a fixed
decomposition would simply 
correspond to a \emph{single} state and thus $W=1$.

According to Boltzmann and Planck, the entropy is obtained by 
\begin{equation}\label{S}
S = k \ln W = k N \ln N - k \sum_i N_i \ln N_i = 
-kN \sum_i \frac{N_i}{N} \ln \frac{N_i}{N}
\end{equation}
for large numbers $N_i$ and the most probable distribution is given by the
maximum of $S$ under the conditions 
\begin{equation}\label{NE}
\sum_i N_i = \sum_i N f(\varepsilon_i)  \left( \tau \Delta \varepsilon
\right)^3 = N, \quad 
\sum_i N_i \,\varepsilon_i = 
\sum_i N f(\varepsilon_i) \left( \tau \Delta \varepsilon \right)^3 
\varepsilon_i = E.
\end{equation}
This procedure superficially resembles the derivation of the canonical
ensemble, however, its spirit is completely different. 
We know that the ST equation is only valid for a dilute gas, 
and Tetrode's derivation implicitly assumes that the occupation
numbers, i.e.\ the numbers of
particles occupying the energy levels of single-particle states, are very
small; otherwise the expression for the number of distinguishable
configurations in phase space would be much more complicated than 
$W'/N!$ and effects of spin and statistics would have to be taken into account. 
However, Sackur in his derivation assumes the opposite, 
namely occupation numbers $N_i \gg 1$.

Finding the maximum of $S$ of equation~(\ref{S}) amounts to computing
the stationary point of the functional $-\int \dd \varepsilon f \ln f$,
under the conditions of a fixed total number of atoms and a fixed
energy, where the function $f$ is defined in the Ansatz~(\ref{ansatz}).
The sought for stationary point is obtained from the maximum of 
\begin{equation}
\Phi(f, \varepsilon) = -f \ln f + \left( \alpha' + 1 \right) f - 
\beta \varepsilon f, 
\end{equation}
where the parameters $\alpha'$ and $\beta$ are Lagrange multipliers:
\begin{equation}
\frac{\partial \Phi}{\partial f} = 
-\ln f + \alpha' - \beta \varepsilon = 0
\quad \Rightarrow \quad 
f(\varepsilon) = e^{\alpha' - \beta \varepsilon} = 
\alpha e^{-\beta \varepsilon}
\quad \mbox{with} \quad \alpha = e^{\alpha'}.
\end{equation}
Eventually, Sackur arrives at the Boltzmann distribution 
\begin{equation}
f(\varepsilon) = \alpha e^{-\beta \varepsilon}.
\end{equation}
Plugging $N_i$ with this $f$ 
into formula~(\ref{S}) and using
equation~(\ref{NE}), the simple expression  
\begin{equation}\label{S2}
S = -3kN \ln (\tau \Delta\varepsilon) - kN \ln \alpha + k \beta E
\end{equation}
for the entropy ensues.

In equation~(\ref{S2}) there are three unknowns: 
$\tau \Delta\varepsilon$, $\alpha$ and $\beta$. At this point, 
referring to Sommerfeld~\cite{sommerfeld}, Sackur 
states that the smallest action that can take place in nature is given by
Planck's constant $h$. Therefore, he makes the bold assumption that
\begin{equation}
\tau \Delta \varepsilon = h,
\end{equation}
which he had already made successfully for the derivation of Planck's
law of radiation in the same paper. 
The other two parameters are in principle determined by
equation~(\ref{NE}). Sackur then argues that, for simplicity, in the
two integrals of equation~(\ref{NE}) summation can be replaced by
integration. For this purpose he makes the following step:
\begin{equation}\label{pk}
\varepsilon_k = \frac{p_k^2}{2m} \;\; (k = x,y,z) 
\quad \Rightarrow \quad \dd \varepsilon_k = \frac{p_k}{m}\, \dd p_k =
  \frac{\bar x_k}{\tau}\, \dd p_k,
\end{equation}
where the $\bar x_k$ are the average Cartesian components
of the distance covered by the atoms during the time $\tau$. 
%Apparently, in view of the summation in equation~(\ref{NE}), 
%the momentum components in equation~(\ref{pk}) are positive,
%therefore, $\bar x_k$ is the average of $|x_k|$ during the
%time interval $\tau$. 
Then Sackur connects the product of the three average
distances with the volume $V$ of the gas by equating it with the
volume per atom:
\begin{equation}\label{v/n}
\bar x \bar y \bar z = \frac{V}{N}.
\end{equation}
It is hard to understand why this equation should hold, but with 
equations~(\ref{pk}) and~(\ref{v/n}) he effectively introduces an
integration $\dd^3 x\, \dd^3 p$ in phase space.\footnote{These
  manipulations introduce an ambiguity in the integration boundaries: In
  $\dd\varepsilon_k$ the integration is from zero to infinity, while
  in $\dd p_k$ Sackur integrates from minus infinity to plus infinity.} 
Moreover, since Sackur nowhere introduces the concept of indistinguishable
atoms, he needs the factor $1/N$ in equation~(\ref{v/n}) for
avoiding Gibbs paradox, as we will see shortly. So he ends up with 
\begin{equation}
\tau^3 \dd \varepsilon_x \dd \varepsilon_y \dd \varepsilon_z = 
\frac{V}{N}\, \dd p_x \dd p_y \dd p_z
\end{equation}
for the integration in equation~(\ref{NE}) and obtains 
\begin{equation}
1 = \frac{\alpha V m^3}{N} \left( \frac{2\pi}{m \beta} \right)^{3/2}
\quad \mbox{and} \quad
E = \frac{3\alpha V m^3}{2\beta} \left( \frac{2\pi}{m \beta}
\right)^{3/2}.
\end{equation}
These two equations are easily solved for $\alpha$ and
$\beta$. Plugging the solution 
\begin{equation}
\beta = \frac{3N}{2E} \quad \mbox{and} \quad
\alpha = \frac{N}{V} \left( \frac{3N}{4\pi mE} \right)^{3/2}
\end{equation}
into equation~(\ref{S2}), Sackur arrives at his final result 
\begin{equation}\label{s-sackur}
S(E,V,N) = kN \left( \frac{3}{2} \ln \frac{E}{N} + 
\ln \frac{V}{N} + \frac{3}{2} \ln \frac{4\pi m}{3 h^2} + 
\frac{3}{2} \right).
\end{equation}

Comparing this expression with Tetrode's result~(\ref{s-tetrode}), 
we see that there is a difference in the last term in parentheses; 
Sackur has $3/2$ while while Tetrode has the correct number $5/2$.
Thus 
\begin{equation}
\left. S(z=1) \right|_\mathrm{Tetrode} - 
\left. S \right|_\mathrm{Sackur} = kN,
\end{equation}
which Sackur observed and commented upon in~\cite{sackur}. It is
interesting to note that in his previous paper~\cite{sackur2} Sackur
actually had the correct number.
It is kind of amazing that Sackur, with his line of reasoning, arrives
at nearly the correct result, being off only by $kN$. This difference
is indeed important for the comparison of the entropy
formula with the data from vapor
pressure of mercury~\cite{tetrode,sackur}; anticipating
equation~(\ref{vp}), we see that a determination of Planck's constant
with Sackur's formula would result in a value which is too low by a
factor of $e^{-1/3} \approx 0.72$ where $e$ is Euler's number.

We conclude this section with a comment on equation~(\ref{v/n}). 
We know that $S$ is an extensive quantity, i.e. 
$S(\zeta E, \zeta V, \zeta N) = \zeta S(E,V,N)$ holds for all
$\zeta > 0$. If the factor $1/N$ had
been absent in equation~(\ref{v/n}), we would have to replace $V$ by
$NV$ in equation~(\ref{s-sackur}); but then $S$ would not be an
extensive quantity, as one can easily check.

%\subsection{The entropy and the canonical partition function}
\subsection{Discussion}
Let us present here, in particular, for comparison with Sackur's treatment, 
the derivation of the entropy of a
monoatomic ideal gas by using the canonical partition function $Z$. 
Since we are dealing with non-interacting particles, $Z$ is given by
\begin{equation}
Z = \frac{Z_1^N}{N!},
\end{equation}
where $Z_1$ is the partition function of a single particle. The factor $1/N!$
is present to take into account that the particles are indistinguishable.
Then the entropy is given by
\begin{equation}\label{SZ1}
S = k \left( \ln Z + \beta E \right) =
kN \left( \ln \frac{Z_1}{N} + 1 + \frac{\beta E}{N} \right),
\end{equation}
where $E$ is the total energy of the $N$ particles and $\beta = 1/(kT)$. 
Furthermore, Stirling's formula has been used to replace 
$\ln N!$ by $N(\ln N - 1)$. If $E/N$ does not depend on $N$, which is the case
for the ideal gas, this equation displays the full dependence on $N$.
For the monoatomic ideal gas, in the classical approximation, the
single-particle partition function is given by the integral 
\begin{equation}
Z_1 = \frac{1}{h^3} \int_{\mathcal{V}} \dd^3x \int \dd^3p\, \exp
\left( -\beta \frac{{\vec p}^{\,2}}{2m} \right) = \frac{V}{\lambda^3}
\quad \mbox{with} \quad
\lambda = \frac{h}{\sqrt{2\pi m kT}}
\end{equation}
being the thermal de Broglie wave length. The integration domain $\mathcal{V}$
is the space taken by the gas, i.e.\ the container with volume
$V$. Plugging $Z_1$ into equation~(\ref{SZ1}) yields the desired entropy 
\begin{equation}\label{Scan}
S(T,V,N) = kN \left( \ln \frac{V}{\lambda^3N} + \frac{5}{2} \right)
\end{equation}
as a function of temperature, volume and particle number.

We compare Tetrode's and Sackur's result with the entropy formula~(\ref{Scan})
by substituting
\begin{equation}
E = \frac{3}{2}\,NkT
\end{equation}
in equations~(\ref{s-tetrode}) and~(\ref{s-sackur}).\footnote{In Tetrode's
  formula we set $z=1$.} We find what we have
announced earlier: Tetrode's result exactly agrees with equation~(\ref{Scan}),
while Sackur's result differs by $kN$. We can easily locate the origin of the
difference. Considering the definitions of $\alpha$ and $Z_1$ and taking into
account equation~(\ref{NE}), we find that
\begin{equation}
\alpha = \frac{N}{h^3 Z_1}.
\end{equation}
Insertion of this expression into equation~(\ref{S2}) leads to the
entropy~(\ref{SZ1}), with the ``1'' within the parentheses being absent.
%Sackur is missing the second term in equation~(\ref{SZ1}), which
%stems from $-\ln N! \simeq N(-\ln N + 1)$. 
Effectively Sackur replaces $\ln N! \simeq N(\ln N - 1)$ by $N\ln N$ 
in his derivation and does, therefore, not fully
take into account indistinguishability of the atoms.

The entropy of the monoatomic ideal gas as a function of the pressure $p$
instead of the volume $V$ is obtained with the ideal-gas equation by the
substitution $V = NkT/p$.

As mentioned in the introduction, Sackur and Tetrode tested their equation on
mercury vapor. This element has seven stable isotopes with various nuclear
spins $s_k$~\cite{aw}. Therefore, in principle for mercury one has to add the 
corresponding residual entropy 
\begin{equation}\label{Sres}
S_\mathrm{res}(\mbox{Hg}) = Nk\,\sum_{k=1}^7 P_k 
\left( -\ln P_k + \ln(2s_k + 1) \right),
\end{equation}
where the $P_k$ are the isotopic abundances ($\sum_k P_k = 1$), to the
Sackur--Tetrode formula. Of course, in~1912 the mercury isotopes were not
known. However, as we will see in the next section, in the mercury
test only the entropy difference between gaseous and liquid phases is
relevant. For both phases, however, the same residual entropy is
expected and thus 
$S_\mathrm{res}(\mbox{Hg})$ of equation~(\ref{Sres}) drops out.

\section{The vapor pressure of mercury and Planck's constant}
\label{vapor pressure}
How to subject the \emph{absolute} entropy of a monoatomic ideal gas
to experimental scrutiny? Sackur and Tetrode applied the following procedure. 
Consider the latent heat $L(T)$ of a monoatomic substance for the phase
transition from the liquid to the gaseous phase. 
In terms of the absolute molar entropies, the latent heat is given by
\begin{equation}\label{L}
L(T) = T \left( 
s_\mathrm{vapor}(T, \bar p(T)) - s_\mathrm{liquid}(T, \bar p(T)) \right),
\end{equation} 
where $\bar p(T)$ denotes the pressure along the coexistence curve,
i.e.\ the vapor pressure.
If the vapor behaves in good approximation like a monoatomic ideal gas,
then the Sackur--Tetrode equation in the form
\begin{equation}\label{st-molar}
s_\mathrm{vapor} = R \left( \ln \frac{kT}{\bar p \lambda^3} + \frac{5}{2} 
\right)
\end{equation}
with the molar gas constant $R$ 
can be substituted for $s_\mathrm{vapor}(T, \bar p(T))$.
For the liquid phase, neglecting the $p$-dependence, the absolute entropy can
be expressed as an integral over the heat capacity:
\begin{equation}\label{s-liquid}
s_\mathrm{liquid} = \int_0^T \dd T'\, \frac{c_p(T')}{T'}.
\end{equation}
Note that here the integration includes the 
solid and liquid phases, and the latent heat of melting. 
After insertion of 
$s_\mathrm{vapor}$ and $s_\mathrm{liquid}$ into
equation~(\ref{L}), one obtains an expression for the vapor pressure:
\begin{equation}\label{vp}
\ln \bar p(T) = -\frac{L(T)}{RT} + 
\ln \frac{(2\pi m)^{3/2} (kT)^{5/2}}{h^3} + \frac{5}{2} - 
\int_0^T \dd T'\, \frac{c_p(T')}{RT'}.
\end{equation}
Similar derivations can be found in~\cite{zemanski,reif}.
Since equation~(\ref{vp}) is a direct consequence of
equation~(\ref{st-molar}), it serves as a testing ground for the
Sackur--Tetrode equation. For this test not only data 
on the vapor pressure $\bar p(T)$ are needed, 
but also data on the latent heat $L(T)$ and the heat
capacity $c_p(T)$ in the condensed phase must be available.
While for $\bar p(T)$ and $L(T)$ it is sufficient to have data in a certain
temperature interval, one needs to know $c_p(T)$ as a function
of $T$ down to absolute zero. In 1912 the most comprehensive set of data was
available on mercury. This was utilized by Sackur and Tetrode to test their
equation. In this test they followed slightly different
approaches. Both employed the value of Planck's constant $h$ as
determined from black-body radiation and inserted it into
equation~(\ref{vp}). Then Sackur directly computed the vapor pressure of
mercury from equation~(\ref{vp}) and compared his results with the
experimental data, whereas Tetrode replaced $h$ in equation~(\ref{vp})
by $zh$ and carried out a fit of $z$ to the data.

Now we want to delineate how Sackur and Tetrode actually performed the
numerical evaluation of equation~(\ref{vp}). 
We follow the exposition of Sackur in~\cite{sackur} because his account is
sufficiently detailed and easy to follow. On the right-hand side of
equation~(\ref{vp}) we have to discuss the term with $L(T)$ and the integral.
In treating the latent heat as a function of
$T$, Sackur uses 
Kirchhoff's equation---see equation~(\ref{kirchhoff}) in the appendix. 
Furthermore, he assumes that in the temperature interval he considers,
which is from $0^\circ\,\mbox{C}$ to $360^\circ\,\mbox{C}$, the 
heat capacity in the liquid phase can be regarded to have the constant value 
$c_p^\mathrm{liquid}$. If at a reference temperature $T_1$ the latent heat is
$L_1$, then due to Kirchhoff's equation
\begin{equation}\label{L1}
L(T) = L_1 + \left(\frac{5}{2}\,R  - c_p^\mathrm{liquid} \right) (T-T_1).
\end{equation}
The integral on the right-handed side of equation~(\ref{vp}) is treated by
splitting it into the part in the solid phase, the contribution of the
phase transition, and the part in the liquid phase. Denoting the 
latent heat of melting by $L_m$ and the melting point by $T_m$, this
integral reads
\begin{equation}
\int_0^T \dd T'\, \frac{c_p(T')}{T'} = 
\int_0^{T_m} \dd T'\, \frac{c_p^\mathrm{solid}(T')}{T'} + 
\frac{L_m(T_m)}{T_m} +
c_p^\mathrm{liquid} \ln \frac{T}{T_m}.
\end{equation}
Again the approximation that the heat capacity of the liquid is 
temperature-independent has been used. 
Implicitly the additional approximation that the
melting temperature $T_m$ is independent of the pressure has been made.
The final form of the vapor pressure, 
prepared for the numerical evaluation, is thus 
\begin{eqnarray}
\ln \bar p(T) &=& - 
\frac{L_1 + \left( c_p^\mathrm{liquid} - \frac{5}{2}R \right)T_1}{RT}
+ \frac{5}{2} \ln T - \int_0^{T_m} \dd T'\, \frac{c_p^\mathrm{solid}(T')}{RT'}
\nonumber \\[2mm] && \label{vp1}
-\frac{L_m(T_m)}{RT_m} - \frac{c_p^\mathrm{liquid}}{R} \ln \frac{T}{T_m} 
+ \ln \frac{(2\pi m)^{3/2} k^{5/2}}{h^3} + \frac{c_p^\mathrm{liquid}}{R}.
\end{eqnarray}
This equation corresponds to Sackur's equation on top of p.~82
of~\cite{sackur} and we have written the terms in the same order as
there. We have refrained, however, from converting the natural logarithm to
the logarithm to the base of ten, which was used by Sackur. 
As mentioned earlier, Sackur and Tetrode actually determine 
the \emph{chemical constant}, defined as
\begin{equation}\label{chem}
\mathcal{C} = \frac{1}{\ln 10} \times \ln \frac{(2\pi m)^{3/2} k^{5/2}}{h^3} = 
\log \frac{(2\pi m)^{3/2} k^{5/2}}{h^3},
\end{equation}
from the data and compare this value of $\mathcal{C}$ with the value computed
with Planck's constant obtained from black-body radiation.
At that time, the chemical constant was a commonly used quantity. 
It appears not only in the vapor pressure
but also in the law of mass action of chemical reactions in the gas
phase~\cite{nernst}. Note that the conversion of the logarithm
mentioned above brings about a 
division by $\ln 10 \approx 2.3026$ in many places in the equations
in~\cite{tetrode,sackur}. 

In equation~(\ref{vp1}), in the integral over 
$c_p^\mathrm{solid}(T)/T$ both Sackur and Tetrode use a model by
Nernst~\cite{nernst1} for the specific heat of solid mercury. This
model is a kind of Einstein model~\cite{einstein} but is sums two
frequencies, $\omega$ and $2\omega$. 
It is interesting to note that the paper of Debye concerning the Debye
model~\cite{debye} has a ``received date'' July 24, 1912, and is thus
prior to Sackur's paper~\cite{sackur}. Actually, Sackur refers to it
in~\cite{sackur}, but only in the part concerning Planck's law of
radiation; in the integration over the solid phase of mercury he uses
nevertheless Nernst's model.

We conclude this section by summarizing and commenting on the
approximations which lead to equation~(\ref{vp1}). In essence the
following approximations have been made:
\begin{enumerate}
\renewcommand{\theenumi}{\roman{enumi}}
\item
The vapor is treated as a classical ideal gas.
\item
The molar volume $v_l$ of the liquid is neglected compared to the
molar volume $v_g$ of the vapor.
\item
In the liquid phase the dependence on $p$ of the isobaric heat capacity is 
negligible in the considered temperature interval.
\item\label{4}
There are two technical assumptions which facilitate the numerics: 
The temperature dependence of the heat capacity in the liquid phase is 
neglected and the melting temperature $T_m$ is pressure independent.
\end{enumerate}
From the first assumption it follows that the heat capacity of a
monoatomic vapor is constant with the value 
\begin{equation}\label{cp-v}
c^\mathrm{vapor}_p = \frac{5}{2}\,R,
\end{equation}
which is an important ingredient in equation~(\ref{L1}). 
The thermal equation of state, 
\begin{equation}
p V = n_m R T,
\end{equation}
where $n_m$ the number of moles of the gas, has been used in
equation~(\ref{st-molar}) and in the derivation of
Kirchhoff's equation---see appendix.
The second assumption, which occurs only in the derivation of
Kirchhoff's equation, is well justified because the order of magnitude
of the ratio of the molar volumes is $v_g/v_l \sim 10^3$. 
To discuss the third assumption we note that via the Gibbs potential we
obtain the relation
\begin{equation}
\left. \frac{\partial c^\mathrm{liquid}_p}{\partial p} \right|_T = 
-T v \left( \alpha^2 + \left. \frac{\partial \alpha}{\partial T} \right|_p
\right) \quad \mbox{with} \quad 
\alpha = \frac{1}{v} \left. \frac{\partial v}{\partial T} \right|_p,
\end{equation}
where $\alpha$ is the thermal expansion coefficient. This equation
leads to a linear approximation of the heat capacity with respect to the
pressure:
\begin{equation}
c^\mathrm{liquid}_p(T,p) \approx c^\mathrm{liquid}_p(T,p_0) - 
T \left( \alpha^2 + 
\left. \frac{\partial \alpha}{\partial T} \right|_p \right)_{p=p_0}
v(T,p_0)\,(p-p_0).
\end{equation}
The pressure $p_0$ is a reference pressure. 
It is well known that the $p$-dependence of $c_p$ for liquids is
suppressed for two reasons. First of all, the product 
$v p \sim 1\,\mbox{J}\, \mbox{mol}^{-1}$ where $v$ is the molar volume
of the liquid and $p \sim 1\,\mbox{bar}$ is rather small. Secondly, 
the thermal expansion coefficient $\alpha$ of a liquid is small as
well; for instance for mercury
$\alpha \approx 1.8 \times 10^{-4}$\,K$^{-1}$ at 1\,bar. Thus, the
third assumption is very well justified. However, in general 
the heat capacity of a liquid depends on the temperature, although not
drastically. For mercury it drops by $4\%$ between 
$-38.84^\circ\,\mbox{C}$, which is the melting point, 
and $200^\circ\,\mbox{C}$~\cite{crc1}.

\section{Our fit of Planck's constant to mercury data}
\label{fit}
It is worthwhile to use the thermodynamic data on mercury available at present
and employ a slight variation of the method of Sackur and Tetrode described in
the previous section in order to check the accuracy with which Planck's
constant can be determined in this way. 
We follow Tetrode's approach in replacing $h$ by $zh$ in
equation~(\ref{vp}). In the following we will plug in the modern meanvalue of
$h$ and determine $z$ from the data. The best 
modern value of $h$, recommended by CODATA~\cite{nist-fc}, is 
\begin{equation}\label{h}
6.626 069 57(29) \times 10^{-34}\,\mathrm{J}\,\mathrm{s}
\end{equation}

In order to account for the slight temperature dependence of the heat capacity
of liquid mercury we make the ansatz
\begin{equation}\label{cp-l}
c_p^\mathrm{liquid}(T) = a_0 + a_1 T + a_2 T^2
\end{equation}
and fit the coefficients $a_0$, $a_1$ and $a_2$ to the input data from the
table presented in~\cite{crc1}. In this table one can also read off that from
the melting point up to a temperature of about $200^\circ\,\mbox{C}$ 
the heat capacity of gaseous mercury agrees exactly with the ideal-gas
value~(\ref{cp-v}). Thus we confine ourselves to the temperature interval from 
$-38.84^\circ\,\mbox{C}$ to $200^\circ\,\mbox{C}$, in which the
ansatz~(\ref{cp-l}) should be sufficient. 
With equations~(\ref{L}) and~(\ref{cp-l}), and taking into account Kirchhoff's
equation, we obtain
\begin{equation}
L(T) = L_0 + \frac{5}{2}\,R (T-T_0) - a_0 \left( T-T_0 \right) -\frac{1}{2} a_1
\left( T^2 - T_0^2 \right) - \frac{1}{3} a_2 \left( T^3 - T_0^3 \right),
\end{equation}
while inserting equation~(\ref{cp-l}) into the entropy
formula~(\ref{s-liquid}) gives
\begin{equation}
s_\mathrm{liquid}(T) = s_0 + a_0 \ln \frac{T}{T_0} + a_1 (T-T_0) + 
\frac{1}{2} a_2 \left( T^2 - T_0^2 \right).
\end{equation}
As a reference temperature we take $T_0 = 298.15\,\mbox{K}$, which allows us
to use the enthalpy of formation and the standard molar entropy from the CODATA
Key Values for Thermodynamics~\cite{key}:
\begin{equation}
L_0 = 61.38 \pm 0.04\,\, \mbox{kJ}\,\mbox{mol}^{-1}, \quad
s_0 = 75.90 \pm 0.12\,\, \mbox{J}\,\mbox{K}^{-1}\,\mbox{mol}^{-1}.
\end{equation}
The value of $s_0$ saves us from the non-trivial task of determining the
integral in equation~(\ref{s-liquid}) with the boundaries $T=0$ and $T=T_0$.

The input data for the vapor pressure of mercury we take from the table
in~\cite{crc2}. In the legend of this table estimated uncertainties of the
vapor pressure values are given, which we use in the method of least squares in
order to fit the parameter $z$. 
A further input parameter is the atomic weight of mercury,
$A = 200.95(2)$~\cite{aw}. 
The mass value for mercury is then $m = Au$ where $u$ is the
atomic mass unit. For the determination of $h$ from mercury data we can safely
neglect errors in the physical constants $R$, $k$ and $u$.

With the above input, 
our best fit value for $z$ is $\bar z = 1.003$ at $\chi^2_\mathrm{min} = 4.2$. 
Since we have at disposal vapor pressure measurements at 75
temperatures~\cite{crc2} in the 
considered interval, but we determine only one
parameter, the number of degrees of freedom is 74. 
For such a large number of degrees
of freedom the above value of the minimal $\chi^2$ tells us that the fit is
perfect. We take into account the following sources of uncertainties in $z$:
the statistical error determined by $\chi^2(z) = \chi^2_\mathrm{min} + 1$,
the errors in $A$, $L_0$ and $s_0$, and an error in $c_p$. 
We obtain the uncertainties $\pm 0.0002$ for the statistical error and 
$\pm 0.0005$ for the error in $A$. These errors are one order of magnitude
smaller than the errors originating in $L_0$ and $s_0$ which are 
$\pm 0.004$ and $\pm 0.005$, respectively. We have no information on the error
in the heat capacity of liquid mercury in~\cite{crc1}. Therefore, we simply
vary $a_0$ by $\pm 1\%$ as a generous error estimate~\cite{giauque}; the
resulting uncertainty, however, is smaller than the statistical error. In
summary, our value of $z$ is 
\begin{equation}
z = 1.003 \pm 0.004\,(L_0) \pm 0.005\,(s_0).
\end{equation}
Of course, the error estimate above is not a sound statistical computation,
but we can safely argue that, with existing thermodynamic data on the
equilibrium of liquid and gaseous phases of mercury, Planck's constant can be
determined with an accuracy of about one percent. Improving the accuracy of
$L_0$ and $s_0$ might improve the determination of $h$, but due to the
approximations pointed out in the previous section, 
thermodynamic data can most probably never compete with quantum physics data
for this purpose.

\section{Conclusions}
\label{conclusions}
Planck's quantum hypothesis in 1900 was a revolutionary step 
which he justified by referring to Boltzmann, because in this way he could
count the number of different photon states and compute the entropy of
a photon gas by using formula~(\ref{SW}).
The importance of the quantum hypothesis became clear only gradually. 
In the beginning, Planck's constant played a role in loosely connected
or seemingly unconnected phenomena. The unified perspective was achieved
only later with quantum mechanics and quantum field theory. However, 
the importance of the quantum hypothesis for atomic and
molecular physics, including thermodynamic quantities like heat
capacities, was suspected quite early, for instance, by
Sommerfeld~\cite{sommerfeld} who connected Planck's constant
with the ``action\footnote{Here action has the usual meaning of the
  time integral over the Lagrangian.} in pure molecular processes.'' 

In the beginning, apart from black-body radiation, the phenomena to
which the quantum hypothesis could be applied were scarce. 
In 1905 it was used by Einstein to explain the photoelectric effect. 
A bit later Johannes Stark could interpret  
features of the light spectrum emitted by canal rays and of the X-ray
spectrum produced by the impact of electrons with the help of the
quantum hypothesis. In 1907 Einstein put forward the ``Einstein
model'' of the heat capacity of solids where $h \nu$ was now
associated with the energy of vibrations of a crystal; this
theory could account for deviations from the 
Dulong--Petit law at high temperatures but gave 
the wrong behavior at low temperatures. This flaw was cured by 
Debye~\cite{debye}, who developed his model practically at the same
time as Sackur and Tetrode derived their equation. 
The Bohr model of the atom was to follow in 1913. 
As a side remark, Ernest Rutherford's paper
on the atomic nucleus appeared in 1911, in the same year when Heike
Kamerlingh Onnes discovered superconductivity. For an extensive
account of the evolution of the ``old quantum theory'' we refer the
reader to~\cite{rechenberg}.

Just as Planck more than ten years earlier, Sackur and Tetrode referred to
Boltzmann in the derivation of their equation. 
One can view the Sackur--Tetrode equation and its successful test with
thermodynamic data as one of the very first confirmations
of Planck's quantum hypothesis. 
This equation was a quite fundamental step towards modern physics as it
demonstrated the ubiquity of Planck's constant in statistical physics. 
We stress once more that the outstanding feature of the papers of Sackur
and Tetrode was the combination of theoretical ideas with an
ingenious usage of experimental data.

One may speculate why the work of Sackur and Tetrode is not that well
known in the physics community as one would expect from its importance
in the development of quantum theory and statistical physics.  
One reason is certainly that both died rather young. 
Sackur (1880--1914), who was actually a
physical chemist, died in an explosion in the laboratory of Fritz
Haber, only two years after the Sackur--Tetrode equation. On the
other hand, Tetrode (1895--1931) was a wunderkind who published his
first research paper, namely the paper on the Sackur--Tetrode
equation, at the age of 17. Later on he rather lived in 
seclusion, though he did publish a few papers which were appreciated
by the community\footnote{Tetrode published a total of six
  papers~\cite{dieks}.} and kept some contact with eminent contemporary 
physicists before he prematurely died of tuberculosis.

\paragraph{Acknowledgements:} The author thanks E.R.\ Oberaigner for
useful discussions and P.O.\ Ludl for a critical reading
of the manuscript.

%%%%%%%%%%%%%%%%%%%%%%%%%%%%%%%%%%%%%%%%%%%%%%%%%%%%%%%%%%%%%%%%%%%%%%%%%%%%

\appendix
\setcounter{equation}{0}
\renewcommand{\theequation}{A\arabic{equation}}

\section{Kirchhoff's equation}

For the transition from the liquid (or solid) to the gaseous phase, 
this equation relates the slope of the latent-heat
curve to the difference of the heat capacities across the
coexistence curve. We derive it here because it is not that well known and
also because we want to give a fairly self-contained account of the
physics around the Sackur--Tetrode equation. 
A derivation of Kirchhoff's equation is also found
in~\cite{zemanski,blundell}.

The starting point is the equation
\begin{equation}\label{Lg}
L(T) = T \left( s_2(T, \bar p(T)) - s_1(T, \bar p(T)) \right)
\end{equation}
for the molar latent heat of the transition from phase~1 to phase~2.
In order to simplify the notation we define
\begin{equation}
\Delta s(T) = s_2(T, \bar p(T)) - s_1(T, \bar p(T)),
\end{equation}
and analogously $\Delta v$ and $\Delta c_p$, the molar volume and heat
capacity differences, respectively, along the coexistence curve. 
Taking the derivative of equation~(\ref{Lg}) with respect to $T$, 
we obtain
\begin{equation}\label{dL}
\frac{\dd L}{\dd T} = \Delta s + 
T \left. \frac{\partial \Delta s}{\partial T} \right|_p + 
T \left. \frac{\partial \Delta s}{\partial p} \right|_T 
\frac{\dd \bar p}{\dd T}.
\end{equation}
Though along the coexistence curve the entropy difference is a function of 
the temperature alone because the pressure along this curve is 
given by $p = \bar p(T)$, 
the partial derivatives in equation~(\ref{dL}) refer to the original
dependence of the entropy on temperature and pressure; of course, after
performing the derivatives, $p$ has to be replaced by $\bar p(T)$.
Next we perform three substitutions in equation~(\ref{dL}). 
Firstly, we note that the molar heat capacity is given by 
\begin{equation}
c_p = T \left. \frac{\partial s}{\partial T} \right|_p.
\end{equation}
Secondly, we use the Maxwell relation
\begin{equation}
\left. \frac{\partial s}{\partial p} \right|_T = 
-\left. \frac{\partial v}{\partial T} \right|_p.
\end{equation}
Thirdly, we apply the Clausius--Clapeyron equation 
\begin{equation}\label{CC}
\frac{\dd \bar p}{\dd T} = \frac{L}{T \Delta v}.
\end{equation}
With these substitutions equation~(\ref{dL}) reads
\begin{equation}\label{ddL}
\frac{\dd L}{\dd T} = \frac{L}{T} + \Delta c_p - 
\left. \frac{1}{\Delta v} \frac{\partial \Delta v}{\partial T}  
\right|_p L.
\end{equation}
So far this equation is general for a phase transition of first order. 
Now we argue that, in the case of the vapor pressure over a liquid (or
solid), the third term on the right-hand side of equation~(\ref{ddL})
cancels the first term to a very good approximation. 
To this end we consider the thermal expansion coefficient defined by
\begin{equation}
\alpha = \frac{1}{v} \left. \frac{\partial v}{\partial T} \right|_p
\end{equation}
which, for an ideal gas, is simply $1/T$. Then, using $v_l/v_g \ll 1$,
we derive 
\begin{equation}
\left. \frac{1}{\Delta v} \frac{\partial \Delta v}{\partial T}  
\right|_p = \frac{\alpha_g - \dfrac{v_l}{v_g} \alpha_l}%
{1 - \dfrac{v_l}{v_g}} \approx \alpha_g \approx \frac{1}{T},
\end{equation}
which proves the cancellation announced above.\footnote{Note that for a liquid
  far below the critical point usually not only $v_l \ll v_g$ holds but also
  $\alpha_l \ll \alpha_g$.} 
With this step we finally end up with Kirchhoff's equation for the latent heat
of vaporization: 
\begin{equation}\label{kirchhoff}
\frac{\dd L}{\dd T} \approx \Delta c_p.
\end{equation}

\end{document}